\documentclass[preprint2]{aastex}

\shorttitle{EUV Emission from Abell 4059}
\shortauthors{Bergh\"ofer, Bowyer \& Korpela}

\begin{document}

\title{EXTREME ULTRAVIOLET EMISSION FROM ABELL 4059
}

\author{Thomas W. Bergh\"ofer,\altaffilmark{1,2} Stuart
  Bowyer,\altaffilmark{1} and Eric Korpela\altaffilmark{1}}

\affil{Space Sciences Laboratory, University of California, Berkeley,
        CA 94720-7450, USA}

\altaffiltext{2}{Hamburger Sternwarte, Universit\"at Hamburg, Gojenbergsweg 
112, D-21029 Hamburg, Germany}

\begin{abstract}
\end{abstract}

We present the results of a search for Extreme Ultraviolet emission in A4059, 
a cluster with an X-ray emitting cluster gas. Our analysis of {\it Extreme
Ultraviolet Explorer} (EUVE) Deep
Survey observations of this cluster shows that it is associated
with diffuse EUV emission. Outside the central 2 arcmin radius the entire
EUV emission detected is explained by the low energy tail of the X-ray emitting
gas. Within the central 2 arcmin region of the cluster we find a deficit of 
EUV emission compared to that expected from the X-ray gas. 
This flux deficit is discussed in the context of the cluster's 
cooling flow. 
The results derived for A4059 are compared to EUVE results obtained for
other clusters such as Coma, Virgo, A1795, and A2199. As part of the study we 
have carried out a detailed investigation of the stability of the EUVE 
Deep Survey detector background. Based on long integrations of blank sky 
over 27 months we disprove claims of substantial time
dependent changes in the detector background by R. Lieu and coworkers.
We also show, contrary to the claim of R. Lieu and coworkers, that the images 
obtained
with the detector are independent of the pulse height threshold of the detector
over a substantial range of threshold settings.

\section{Introduction}

A4059 at $\sim300$ Mpc (z = 0.0487) is well studied in the X-ray, radio, 
and optical.  It is relatively compact and is classified as richness class l 
(Schwartz et al., 1991 and references therein).  It is dominated by a central 
CD galaxy, ESO 349-G010, and contains the radio galaxy PKS 2354-35 (Taylor, 
Barton, \& Gee, l994). Taylor et al. (1994) found that the radio galaxy has 
a double lobe structure extending along the major axis of the host galaxy. 
X-ray observations with HEAO I and EXOSAT (Edge \& Stewart 1991; Schwartz 
et al. 1991; Edge, Stewart \& Fabian 1992) showed the cluster has substantial
X-ray cluster emission with a strong cooling flow.  ASCA observations (Ohashi 
1995) found a chemical abundance gradient in the cluster gas around the 
central CD galaxy.  Huang \& Sarazin (1998) analyzed ROSAT, HRI and PSPC X-
ray data on the cluster and found two X-ray holes near the center of the 
cluster emission.

We have studied A4059 in the hopes that it might shed light on the 
underlying source mechanism for EUV emission in clusters of galaxies.  It is 
a particularly interesting cluster to study in this regard because it appears 
to have achieved a relaxed state (Huang \& Sarazin 1998) and has a cooling 
flow.  These characteristics are similar to those in A1795 and A2199
and Bowyer et al. (1999) showed these clusters did not exhibit excess EUV
 emission. On the other hand, A4059 does contain a radio galaxy, and 
Bergh\"ofer et al. (2000) have shown that radio emission may be at least 
indirectly associated with the EUV emission in the Virgo Cluster.  

We report herein our analysis of the EUV emission from A4059. In view of 
recent claims that the EUVE Deep Survey Telescope detector background is 
uncertain since it varies over time, and that the images obtained are highly 
dependent upon the threshold setting employed (Lieu et al. 1999), we have 
examined these issues in detail. Throughout this paper we assume a Hubble 
constant of 50 Km\,s$^{-1}$\,Mpc$^{-1}$ and q(0) = 0.5.

\section{Data and Data Analysis}  

A4059 was observed on two separate occasions. The first observation was 
carried out in June 1998 and provided 39,373\,s of data; the second observation
was made in November 1998 and provided 93,473\,s of data. During the 
observation in June 1998 the cluster center was placed near the boresight of 
the Deep Survey Telescope about two arcmin away from the known dead spot of 
the detector. 

In order to avoid the dead spot and to minimize other detector position 
dependent effects on the detected EUV emission the second observation of A4059
 in November 1998 was carried out at two detector offset positions. During
the first part of the second observation, 48,856\,s of data was obtained 
$\approx 10\arcmin$ to the right of the optical axis; the cluster was then 
observed for another 44,617\,s $\approx 10\arcmin$ to the left of the optical 
axis. A comparison of the cluster's radial EUV profiles obtained during the 
different EUVE observations shows that the effect of the dead spot during the 
June 1998 observation is small and not visible within the statistical error 
bars.

The reduction of the data was carried out with the EUV package built in IRAF. 
We intended to employ the analysis methods described in Bowyer, Bergh\"ofer 
\& Korpela (1999) and Bergh\"ofer, Bowyer \& Korpela (2000) in this work. 
However, several aspects of these procedures have been questioned by Lieu 
et al. (1999). We note that the criticisms of Lieu et al. were made as 
declarative statements, and were unsupported by any analysis. Nonetheless, 
we have examined their points in detail. In specific, Lieu et al. claim that 
the character of the images obtained with the Deep Survey Telescope are 
dependent upon the lower level pulse height cutoff employed, and that the 
cutoff threshold for valid data varies substantially over the face of the 
detector.  In addition, these authors claim that the background for 
observations with the Deep Survey Telescope must be taken nearly 
simultaneously with the observation itself since this background varies over 
time. 

To examine the claim that the pulse-height threshold level setting varies 
widely over the face of the detector, we first assembled 363\,Ks of data from 
a variety of blank fields. We then examined the effect of varying the lower 
level threshold on the summed image of these fields. In specific, we 
constructed an image taken with a lower level setting of 2700  (a detector 
control number that is linearly related to the voltage of the pulse-height 
cutoff setting). We then constructed an image taken with a lower level cutoff 
of 1280.  We compared these two images using a cross correlation analysis and 
found the two had a correlation coefficient of 0.97.  Hence the use of a lower
level threshold cutoff that varies by well over a factor of two has 
essentially no effect on the final images and using any threshold within this 
range provides valid data.

We next studied the claim that the background data must be taken almost 
simultaneously with data from the field to be studied because the stability 
of the Deep Survey Telescope detector sensitivity function varies over time. 
We found that virtually every blank field observation made with the EUVE Deep 
Survey Telescope is well fit with a two-parameter profile.  One profile is a 
flat background that varies from observation to observation (though typically 
by less than a factor of two). This reflects the different charged particle 
backgrounds encountered at different times. Such background changes are
well known from other detectors in space-borne instruments. 
We note, for example, that there are several types
of slowly varying background in the ROSAT PSPC data that are routinely 
accounted for in the analysis of data obtained with that instrument. 

The second profile needed to parameterize a blank field is the detector 
sensitivity variation over the face of the detector. This profile accounts for
telescope vignetting, obscuration by support structures in the detector 
window, and other factors.  To study the stability of this parameter, we added
a number of blank fields, individually obtained over a period of 27 months, to
obtain a total data set of 363\,Ks of blank sky. We subtracted the appropriate
flat background determined from highly obscured regions at the outer most 
parts of the field of view near the filter frames from each of these fields. 
We then convolved the data with a 32 pixel wide Gaussian corresponding to the 
instrument resolution. We show the result in Figure\ 1a. A previous assemblage 
of 425\,Ks of data obtained from a different set of blank fields and processed
in the same manor is shown as Figure 1 of Bowyer et al. (1999) and is 
reproduced here as Figure\ 1b. 
\begin{figure}
  \resizebox{\hsize}{!}{\includegraphics{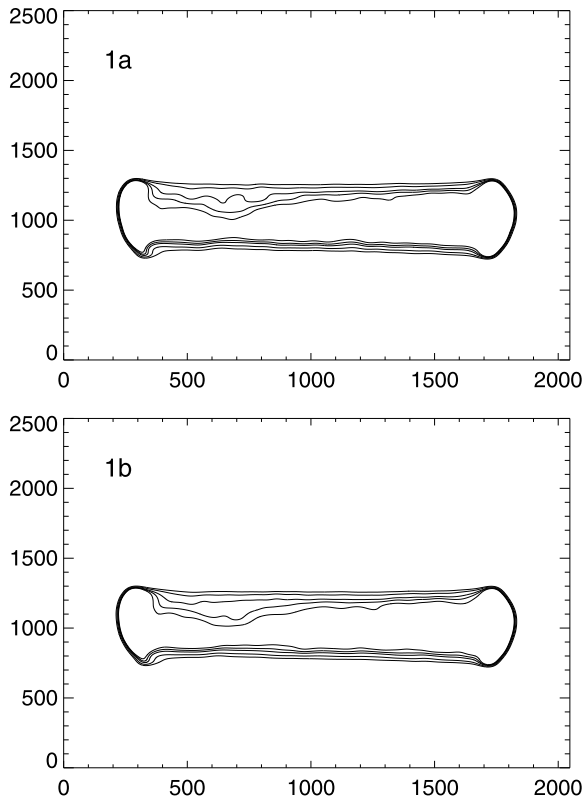}}
  \caption{The EUVE Deep Survey Telescope Detector Sensitivity over
the face of the detector.  Figure 1a shows 363\,Ks of data obtained from a 
number of blank field observations over a period of 27 months. Figure 1b shows
a separate assemblage of 425\,Ks obtained from a different set of blank fields
processed in the same manner. Both plots provide contour levels of 95\%, 90\%,
85\%, 75\%, and 65\%. The two data sets are identical to within the
statistical errors.} 
\end{figure}

We compared these two data sets and found these two images are correlated at 
the 97\% level over the active area of the Deep Survey Telescope detector. 
Given that the statistical errors in each element in each data set are about 
1.5\%, this is consistent with there being no difference between the images. 
We also employed a separate test to check the similarity of these two images. 
We computed the reduced $\chi^2$ of the fit of one image to the other and 
obtained a result of 1.05, again indicating that the two data sets are 
identical to within the statistical uncertainties.

Because this profile is stable there is no need to obtain a contemporaneous 
background for a particular individual observation.  By combining a number of 
blank field observations, this effective detector sensitivity profile can be 
established to any level of desired statistical certainty.  

We note that a similar type of correction for the variation of the detector 
sensitivity over the field of view is routinely used to correct ROSAT PSPC 
observations where the detector sensitivity variations in that instrument are 
incorporated into an effective area exposure map (Snowden et al 1994).  A 
similar procedure is also routinely used in ground based observations of 
astronomical fields with CCD detectors where the procedure is designated as 
``flat fielding''.

In consideration of this demonstration of the validity of our analysis 
procedure, we began our analysis of the EUV data on A4059. We first 
examined the pulse-height spectrum of the detected events to exclude 
background events with pulse-heights significantly larger or lower than the 
source events. Events with low pulse-heights are primarily noise events, and 
events with very large pulse heights are primarily cosmic ray events. The 
photon events are concentrated in a Gaussian profile on top of a 
noise background that rises exponentially towards lower pulse-heights. A
demonstration of this effect is shown in Figure 2.  Here we show a smoothed
pulse height distribution (PHD) of data from a 77\,Ks blank field 
observation.  The dashed line shows the PHD of the non-photonic background for 
data obtained from highly obscured regions of the detector at the outermost 
parts of the field
of view.  The sold line represents the PHD of events obtained over the entire
portion of the detector exposed to the celestial sky.
The curves have been normalized to the detector area of the samples.  The
difference between the two curves represents the PHD of
photon events. In the case of the June 1998 A4059 data set we excluded 
events with pulse heights below 2,500 and above 12,500. For the November 
observation we excluded events below 2,500 and above 13,500. 
\begin{figure}
  \resizebox{\hsize}{!}{\includegraphics{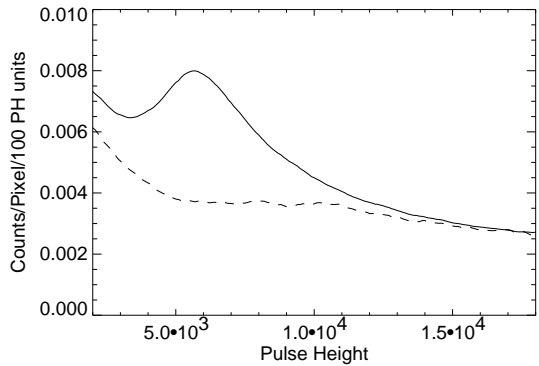}}
  \caption{The pulse height distribution of background and photon
events in the Deep Survey Telescope detector. The dashed line shows the 
non-photonic background of the detector obtained from regions in which 
celestial EUV photons are obscured.  The sold line shows the distribution
over the unobscured portion of the detector that views the sky.  The 
difference between the two represents the pulse height distribution of 
photon events.} 
\end{figure}

Corrections for the detector dead time and telemetry limitations were then 
applied and raw DS EUV images were produced for the observations at the three 
different detector positions. For each of these images a flat non-photonic 
background, determined from highly obscured regions at the outer most parts of
the field of view near the filter frames as discussed above, was subtracted 
from the image. 
We computed 
azimuthally averaged radial emission profiles of A4059 centered on 
the respective detector positions of the cluster  from the 
individual EUV Deep Survey images.

The detector sensitivity map (or flat field) that we constructed from the 
assemblage of 788\,Ks blank field observations was then used to determine 
radial sensitivity profiles for the three detector positions where the cluster
was centered during the observations. These radial sensitivity profiles were 
then fit to the respective profiles of the distinct cluster observations at
radii between 15 -- 20\arcmin. In all three cases, the June 1998 observation 
and the two parts of the November 1998 observation, the fit of the sensitivity
map to the observations provide an excellent representation of the data at 
radii larger than $\approx 5\arcmin$.

In Figure\ 3 we show the combined observed azimuthally averaged, radial EUV 
emission profile of the cluster A4059. The background and its statistical
error is shown as a gray shaded area. Here we have used the entire 788\,Ks 
background data set. Given this extensive data set, the statistical 
uncertainties in this background are small. 
\begin{figure}
  \resizebox{\hsize}{!}{\includegraphics{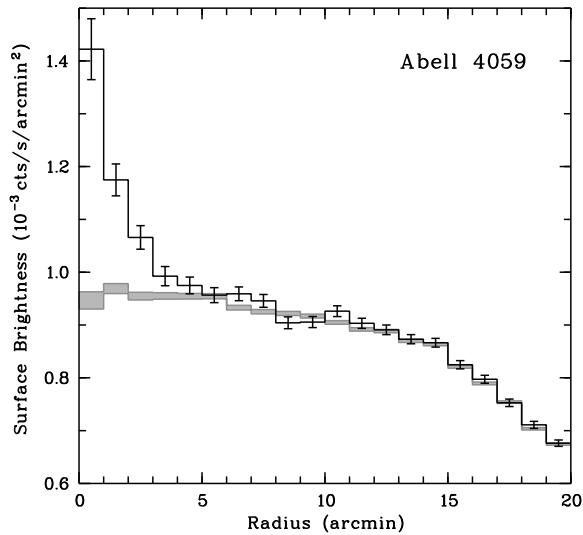}}
  \caption{The azimuthally averaged radial EUV emission profile
of A4059.  The background obtained from 788\,Ks of background and its
statistical errors are shown as grey shaded regions.} 
\end{figure}

The data in this figure shows diffuse EUV emission in A4059 that extends
to a radius of 4 to 5\arcmin. Beyond this point, the radial emission profile 
matches the detector sensitivity function, showing there is no EUV emission at
larger radii.

We next determined the EUV contribution of the low energy tail of the X-ray 
emitting cluster gas in A4059. To this end we analyzed 5,439\,s of ROSAT 
PSPC archival data on this cluster. Using standard procedures implemented in 
the EXSAS software package, we produced a cluster image in the ROSAT hard 
energy band (PSPC channels 51 to 201) and constructed an azimuthally 
averaged 
radial emission X-ray profile centered on the cluster position. 

We employed the cluster X-ray gas temperature profile derived by Huang \& 
Sarazin (1998), the abundance profile from ASCA measurements (Ohashi 1995),
and the MEKAL plasma code to simulate ROSAT PSPC to EUVE DS counts 
conversion 
factors. The correction for the intervening absorption of the Galactic 
interstellar medium (ISM) was carried out using an interstellar hydrogen 
absorption column of N$_{\rm H} = 1.06 \pm 0.1 \times 10^{20}$cm$^{-2}$ 
(Murphy et al. 2000) and an absorption model including cross sections and 
ionization ratios for the ISM as described in Bowyer, Bergh\"ofer \& Korpela 
(1999). Note that the ISM column employed, based on new radio measurements, 
is lower than that used in the analysis of Huang \& Sarazin (1998). We 
conclude that the effect of the somewhat lower ISM column and the improved ISM
cross sections on the X-ray temperatures derived by Huang \& Sarazin is small.
However, an accurate modeling of the foreground absorption is required in our 
work in order to obtain an accurate conversion from ROSAT PSPC counts into 
EUVE DS counts.

For the range of X-ray temperatures (1.8 to 4.0 keV) and element abundances of
0.29 to 0.63 times the solar value as established for A4059 by ASCA 
observations (Ohashi 1995), we determine the ROSAT PSPC hard band to EUVE 
Deep
Survey counts conversion factor fell between 115 to 130. Employing these values
as limits and using the azimuthally averaged radial X-ray emission profile 
derived from the PSPC hard energy band, we derived upper and lower limits for 
the EUV emission from the X-ray emitting gas in the EUVE Deep Survey band 
pass.

\begin{figure}[h]
  \resizebox{\hsize}{!}{\includegraphics{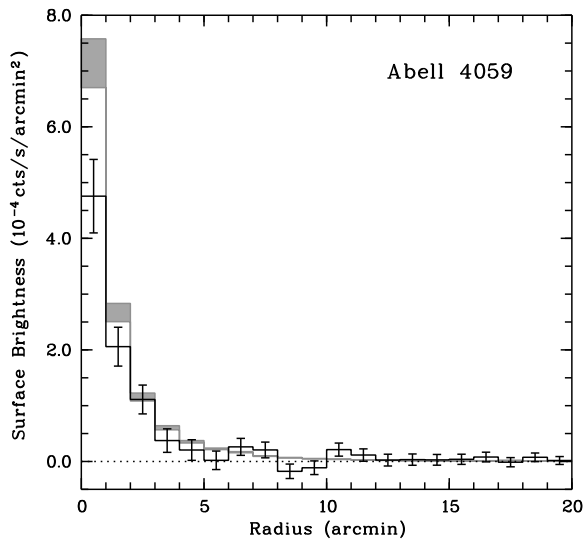}}
  \caption{The EUV emission from the cluster obtained from the data
displayed in Figure 3.  The statistical errors in the background and signal
are added in quadrature.  We also show the EUV emission from the X-ray gas.
There is no evidence of excess EUV emission from the cluster to within the
statistical uncertainties.} 
\end{figure}
In Figure\ 4 we show the EUV emission from the X-ray emitting gas as shaded 
regions with the uncertainties in this emission indicated by the size of the 
shaded region. We also show the EUV emission in the cluster as derived from 
the data displayed in Fig. 3; we have subtracted the background from the 
signal and show the emission with the errors in the signal and background 
added in quadrature. Within the central 2\arcmin\ bins the EUV emission has a 
deficit compared to the expected emission from the diffuse X-ray gas. 
At larger radii the EUV emission is consistent with the EUV contribution of 
the X-ray emitting cluster gas.

\section{Discussion}
\label{discuss}

The data displayed in Figure\ 4 shows there is no excess EUV emission in this 
cluster. There is, in fact, an EUV deficit in the innermost two arc minute 
region of the cluster.  This deficit is similar to the EUV deficit found by 
Bowyer et al. 1999 in the inner core of A1795 and A2199. This effect
is due to absorption by partially ionized material in the cooling flow 
material as has been discussed in Allen et al. (1996) and Bowyer et al. (1999).

Given that A4059 appears to have achieved a relaxed state and has a 
cooling flow, characteristics similar to those in A1795 and A2199, 
which Bowyer et al. (1999) showed did not posses an excess EUV emission, it is
perhaps not surprising that we did not find excess EUV emission in this 
cluster. On the other hand, A4059 does contain a radio galaxy. Since 
Bowyer \& Bergh\"ofer (1998) have shown the EUV excess in the Coma Cluster is 
non-thermal and Bergh\"ofer et al. (2000) have shown that radio emission may 
be at least indirectly associated with the EUV emission in the Virgo Cluster, 
it is reasonable to speculate that EUV emission would be present in this 
cluster as well. Since it is not, there must be some other mechanism involved 
in the production of this flux, either as the underlying cause, or as a 
correlative requirement.

In any attempt to understand the phenomena of EUV excess in clusters of 
galaxies, one must confront the fact that Lieu et al. (1999) claimed to 
have found excess EUV emission from A2199 while Bowyer et al (1999) did 
not. This difference is clearly due to the different analyses procedures 
employed. 

Lieu et al. claim the pulse height threshold varies widely in the EUVE DS 
detector and that this is of fundamental importance. We have discussed this 
issue at length earlier in this paper and show that it is incorrect. Their 
second claim is that a background observation must be made nearly 
contemporaneously with the field observation because of the instability of the
EUVE Deep Survey telescope background.  We discussed this point earlier in 
this paper and show that this point is also incorrect.  

An additional
problem in the Lieu et al. analysis is introduced by the small background 
data set these authors employed. In their Figure\ 4 they claim the radial 
emission profile shows the emission extends out to almost 30\arcmin.  However,
their background observation is about one fifth as long as their cluster 
observation and consequently the errors in this background measurement are 
substantially larger than those for the cluster observation. It is an 
elementary 
statistical requirement that in order to optimize the measurement of a weak 
signal embedded in a background with a similar intensity, both the signal and 
the signal plus the background must have data sets of similar size. This is 
obviously not the case and the large statistical uncertainties in the 
background as shown in Figure\ 4a of Lieu et al. clearly dominate any attempt 
at establishing a signal even out to three arcmin. This can be verified 
by comparing the error bars of the background in Fig 4a with those of the 
signal plus background shown in Figure 4b, although this comparison would be 
easier if the background and its uncertainties were not omitted in Figure\ 4b.

The only difference between our analysis and the analysis of Lieu et al. that 
has any effect is their use of a wavelet analysis to establish their detector 
sensitivity profile or flat field. The profile they obtain using a wavelet 
analysis is 
clearly different than ours as can be seen by comparing Figure\ 1 of Lieu et 
al. showing their background with Figure\ 1 herein showing our profile.  
A wavelet analysis is a complex procedure and may well produce surprising 
results if inappropriately employed. Given the lack of transparency in their 
description of the use of a wavelet analysis, it is impossible to 
determine the source of effects seen in their Figure\ 1.  We note that in an 
extensive anaysis of clusters of galaxies, Mohr et al. (1999) find that 
wavelet analyses of clusters of galaxies by Durret et al (1994), provide 
results that are different from virtually all other authors; they also comment
that it is "difficult to discern the source of the disagreement."
One possibility is 
that the large edge effects seen in the background in our Figure\ 1 are adding
substantial power to their template.  We have asked Lieu (2000, private 
communication) if this has occurred, but he has stated that he will not 
discuss this issue with us.  

An apparent confirmation of a diffuse EUV emission from A2199 is provided
by Kaastra et al. (1999) who use BeppoSAX and EUVE data in their analysis.  
These authors do not describe their reduction of the EUVE data so we cannot 
comment on that part of their analysis. However, these authors claim the 
BeppoSAX data show an excess EUV emission in this cluster. However, in 
analyzing the BeppoSAX data these authors use a curious mix of in-flight 
observational data and ground-based calibration data to determine the detector
sensitivity function. The cosmic X-ray and particle background was obtained 
from in-flight data of empty files taken at high Galactic latitude.  However, 
this in-flight data was not used to determine the effective area, point-spread
function, strong back obscuration and vignetting of the detector. Instead 
these authors use a function that is based on response matrices derived from 
ground-based calibrations and ray trace codes. The lack of transparency of 
this procedure makes it difficult for an outside person to establish the 
validity of this process. We do note, however, that it was the use of a 
ground-based detector response function for the background of the EUVE 
detector that resulted in the errors in the original reports of EUV excesses 
in clusters of galaxies. It was only when in-flight background data were 
employed that the true extent of the EUV excess could be established.    

It is superficially curious that excess EUV emission has been found in the 
Virgo Cluster by Lieu et al. (1996a) and Bergh\"ofer et al. (2000), and in 
the Coma Cluster by Lieu et al. (1996b) and Bowyer et al. (1999) using 
different methods of analysis. Upon reflection it is clear that this is 
because the emission in both these clusters is sufficiently intense and 
extended that both groups obtain clear evidence for an excess. Nonetheless, 
the results obtained using these different methods differ in detail.

While there is agreement that there is an EUV excess in both Virgo and Coma, 
this emission is more complex than previously imagined. In the Virgo Cluster 
EUV emission is associated with the core and jet of the central galaxy M87.  
Additionally, in the vicinity of M87, diffuse emission is observed out to a 
distance of 13\arcmin. The spatial distribution of this flux is incompatible 
with thermal plasma emission originating from a gravitationally bound gas. 
Furthermore, the diffuse EUV emission is not directly correlated with either 
the X-ray or radio emission in the cluster (Bergh\"ofer et al. 2000).  

New data on the Coma Cluster shows the EUV emission is not only associated 
with the main cluster but is also present in the subcluster to the northwest 
(Korpela et al. in progress).  A thermal source for the emission in the 
cluster can be ruled out since the emission is not consistent with a 1/r$^2$ 
gravitational potential.  Nonetheless, the emission is spatially intermixed 
with the high temperature X-ray emitting thermal emission.

\section{Conclusion}

We analyzed unpublished EUVE Deep Survey observations of the cluster of 
galaxies A4059. In order to test the integrity of our results for this
clusters and to test the validity of our background model for the EUVE Deep
Survey instrument we also analyzed 363\,Ks of blank field observations and
compared this to a previous assemblage of 425\,Ks of blank sky.
Using statistical tests of these two backgrounds we 
definitely confirm the stability of the instrument's background and disprove
claims of time dependent changes in the detector background by Lieu et al. 
(1999). Further analysis shows that for a wide range of threshold cutoffs
the pulse-height threshold level setting has essentially no effect on the final
EUVE Deep Survey images, another claim cooked up by Lieu et al. (1999).

Our analysis of the EUVE observations of the cluster of galaxies A4059 
shows that this cluster exhibits diffuse EUV emission. However,
the emission over most of the cluster is that expected from the low energy 
contribution of the X-ray emitting cluster gas. The EUV emission 
in the central 2 arcmin of the cluster shows an EUV deficit. The observed 
level of EUV flux is extremely sensitive to absorption effects and the flux 
deficit demonstrates intrinsic absorption of the X-ray emitting cluster
gas. Together with A1795 and A2199 (Bowyer et al. 1999), A4059 
is the third example of a cluster of galaxies with EUV absorption due to cooler
gas in the existing central cooling flow.

\acknowledgments
This work was supported in part by NASA cooperative agreement NCC5-138 and an
EUVE Guest Observer Mini-Grant.  TWB was supported in part by a Feodor-Lynen 
Fellowship of the Alexander-von-Humboldt-Stiftung.

\end{document}